\documentclass[10pt,sigconf]{acmart}
\usepackage{proj}

\acmYear{2025}\copyrightyear{2025}
\setcopyright{cc}
\setcctype[4.0]{by}
\acmConference[ACM SOSP Poster Session 2026]{ACM SOSP Poster Session 2026}{September 29--October 2, 2026}{Prague, Czechia}

\begin{document}
\title{Rethinking Web Application Firewalls}


\author{Laurin Brandner}
\orcid{0009-0005-8251-9117}
\affiliation{%
  \institution{ETH Zürich}
  \country{}
}

\author{Laurent Vanbever}
\orcid{0000-0003-1455-4381}
\affiliation{%
  \institution{ETH Zürich}
  \country{}
}

\renewcommand{\shortauthors}{Brandner et al.}
\renewcommand\footnotetextcopyrightpermission[1]{}



\maketitle

\section{Introduction}

In recent years, the threat of application-layer (L7) distributed denial-of-service (DDoS) attacks is ever increasing~\cite{online_rapid_reset, online_http2_bomb}. This is problematic due to two reasons: First, while transport-level (L4) attacks rely on huge botnets consisting of tens of thousands of nodes~\cite{antonakakis_understanding, hoque_botnet}, L7 attacks typically have a much higher amplification. This allows attackers to take down a server with only a handful of machines~\cite{online_http2_bomb}. Second, L7 attacks are more difficult to detect because they exploit a flaw in the protocol that makes the peer perform unnecessary work.

To defend against this, network operators deploy web application firewalls (WAFs) such as ModSecurity~\cite{online_modsec}. WAFs are stateful scoring systems that inspect every HTTP request and perform defensive actions if a connection is deemed malicious. Network operators configure them with a \emph{rule set} -- a collection of rules that specify what malicious traffic looks like (e.g. a suspicious HTTP header), and how to handle it (e.g. drop the connection). Each rule takes one or multiple variables as input, performs one or multiple \emph{parse-match-action} operation, and returns new variables as output. At runtime, rules can dynamically reconfigure other rules, and thus have a high degree of interdependence. Typically, a firewall executes all rules for each request. However, e.g. in case of an attack, the firewall can also block traffic with the first matching rule to save resources.

\begin{figure}[t]
    \tikzsetnextfilename{dissect-figure}

\begin{tikzpicture}[
    annot/.style={fill=white, text width=20pt, anchor=east, execute at begin node=\setlength{\baselineskip}{0.75em}},
]
\begin{axis}[
ybar stacked,
ymin=0,
ymax=13,
legend reversed=true,
legend style={
    at={(0.0,1.1)},
    anchor=south west,
    legend columns=-1,
    draw=none,
    /tikz/every even column/.append style={column sep=0.25cm}
},
axis lines=left,
bar width=20pt,
ylabel={Overhead [ms]},
xticklabels={CRS, {$2\times$ CRS}},
xticklabel style={align=center},
xtick=data,
every extra y tick/.style={
        yticklabel style={
            anchor=west,
            fill=white,
            inner sep=0,
            xshift=2.25cm,
        },
},
extra y tick label=\empty,
extra y ticks={5, 10},
extra y tick style={
    grid=major,
    grid style=dashed,
},
enlarge x limits={abs=1.5cm},
height=3cm,
width=\linewidth]

\input{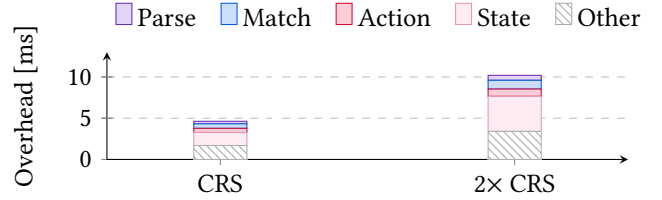}

\end{axis}
\end{tikzpicture}
    \caption{ModSecurity configured with the core rule set (CRS), adds an overhead of 4.6\,ms to each request. This overhead is dominated by the state management.}
    \label{fig:intro:dissect}
\end{figure}

While effective, WAFs are expensive and can slow down a web server significantly. To show this, we deploy an HTTP echo service with ModSecurity, which we configure with the Core Rule Set (CRS)~\cite{online_crs}, the state-of-the-art open-source rule set consisting of approximately 700 rules. We dissect the per-request latency overhead into four categories: The parse, match, and action stage of the rules, along with the state management it needs to coordinate rule reconfiguration and execution. We find that, on average, ModSecurity configured with CRS adds an overhead of 4.6\,ms to each request, making the system more than $6\times$ slower. Duplicating each rule in the CRS (denoted as $2\times$ CRS), adds an overhead of 10.2\,ms (more than $12\times$ slower).This overhead is dominated by state management, as can be seen in \Cref{fig:intro:dissect}.

\paragraph{Related Work} Previous work has addressed this overhead with two techniques. The first technique aims at reducing the overhead of regex matching, e.g. by using more cache-friendly data structures~\cite{choi_dfc} or using SIMD instructions~\cite{wang_hyperscan}. Unfortunately, this approach offers minor performance improvements because regex is not the bottleneck in WAFs. We show this \Cref{fig:intro:dissect}: Regex dominates the matching stage, but matching constitutes only 9.8\% of the overall overhead. The second technique ranks the rules to prioritize rules that match often~\cite{li_accelerating, chen_rulecache}, and blocks traffic with the first match. However, this approach leaves benign traffic unoptimized.

\paragraph{Key Insight} WAF optimization faces a key challenge: WAF rules are extremely expressive to capture the versatility of L7 attacks. This can degrade the performance of the WAF, though: Executing a rule set without preprocessing it first can result in unnecessary work, rendering state management particularly expensive.

Our key insight is that, indeed, we can improve the performance of WAFs without losing versatility. This work proposes \tool, a new WAF architecture that models the interdependency of a rule set as a direct acyclic graph (DAG). This allows it to generate an execution plan that simplifies state management, and enables additional optimizations.

\section{Design}

The following paragraphs describe the key optimizations that \tool employs to jointly accelerate all sources of overhead of state-of-the-art WAF designs.

\begin{figure}[t]
    \tikzsetnextfilename{dag-figure}

\def\innerSep{4pt}
\def\radius{4pt}
\def\pathOffset{0.15cm}
\def\chevronSep{1pt}
\def\containerDistance{0.5cm}
\def\controlPlaneDistance{1cm}
\def\dataPlaneDistance{1cm}

\begin{tikzpicture} [
    x=0.75cm,
    y=0.75cm,
    uchu/.style 2 args={text=uchu-#1-8, draw=uchu-#1-5, fill=uchu-#1-#2},
    dotted/.style={dash pattern=on 0pt off 2\pgflinewidth, line cap=round},
    dashed/.style={dash pattern=on 2\pgflinewidth off 2\pgflinewidth, line cap=round},
    evenlydashed/.style={dash pattern=on 3\pgflinewidth off 3\pgflinewidth, line cap=round},
    solid line/.style={rounded corners=4pt, uchu-gray-8, very thick},
    arrow/.style={solid line, -{Triangle[scale=0.8]}},
    box/.style={rectangle, rounded corners=4pt, minimum height = 0.65cm, minimum width = 0.65cm, inner sep=4, thick, uchu={#1}{1}, draw=uchu-#1-6, minimum height=0.3cm, text depth=0},
    label/.style={inner xsep=4pt, align=center, text depth=0pt},
]

    \node[box={orange}, dashed] (agent var) {\scriptsize\ttfamily user-agent};
    \node[box={orange}, dashed, below=2 of agent var.east, anchor=east] (path var) {\scriptsize\ttfamily path};
    \node[box={orange}, dashed, below=1 of path var.east, anchor=east] (cookie var) {\scriptsize\ttfamily cookie};

    \node[box={blue}, right=1 of path var] (decode transform a) {\scriptsize\ttfamily url-decode};
    \node[box={blue}, right=1 of agent var] (decode transform b) {\scriptsize\ttfamily url-decode};
    \node[box={orange}, dashed, below=1 of decode transform b.west, anchor=west] (pl var) {\scriptsize\ttfamily paranoia level};
    \node[box={blue}, right=1 of cookie var] (re op b) {\scriptsize\ttfamily regex};
    \node[box={blue}, right=1 of pl var] (gt op) {\scriptsize\ttfamily greater than};
    \node[box={blue}, right=1 of gt op] (block action a) {\scriptsize\ttfamily block};

    \node[box={blue}, right=1 of decode transform b] (re op a) {\scriptsize\ttfamily regex};
    \node[box={blue}, right=1 of decode transform a] (re op c) {\scriptsize\ttfamily regex};
    \node[box={blue}, right=1 of re op b] (block action b) {\scriptsize\ttfamily block};

    \node[box={blue}, right=1 of re op a] (log action a) {\scriptsize\ttfamily log};
    \node[box={blue}, right=1 of re op c] (log action b) {\scriptsize\ttfamily log};

    \draw[arrow] (path var) -- (decode transform a);
    \draw[arrow] (agent var) -- (decode transform b);
    \draw[arrow] (decode transform a) -- (re op c);
    \draw[arrow] (decode transform b) -- (re op a);
    \draw[arrow] (re op a) -- (log action a);
    \draw[arrow] (re op c) -- (log action b);
    \draw[arrow] (re op a) -- (pl var);
    \draw[arrow] (re op c) -- (pl var);
    \draw[arrow] (cookie var) -- (re op b);
    \draw[arrow] (pl var) -- (gt op);
    \draw[arrow] (gt op) -- (block action a);
    \draw[arrow] (re op b) -- (block action b);

\end{tikzpicture}
    \caption{\tool models rules as a directed acyclic graph (DAG) where nodes correspond to variables (orange) or operations (blue), and edges to dependencies.}
    \label{fig:design:dag}
\end{figure}
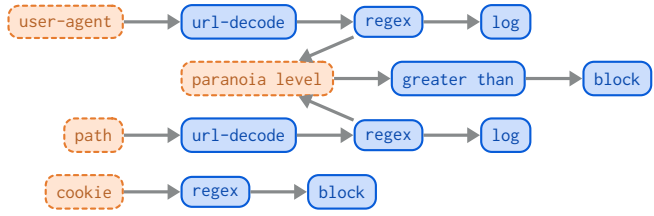

\paragraph{Modelling Rule Interdependence} \tool models the interdependence of the WAF rule set with a DAG, allowing it to manage state more efficiently. The DAG consists of nodes that represent a variable or a \emph{parse}, \emph{match}, or \emph{action} operation, and edges that represent their dependencies. We visualize this in \Cref{fig:design:dag}. The pipeline starts with HTTP headers as the input variables (orange) that go through a series of operations (blue), which optionally return other variables as output. The execution of the pipeline starts by traversing the DAG with the initial variables. Parse operations, e.g. \texttt{url-decode}, transform the variable. Match operations, e.g. \texttt{regex}, conditionally terminate execution of the downstream subgraph. Action operations, e.g. \texttt{block}, perform a side effect and can terminate the entire execution.

During deployment, \tool uses this model to generate an efficient execution plan for the configured rule set. The following paragraphs discuss further optimizations that \tool applies on top of it.

\paragraph{Merging Nodes} \tool merges nodes to produce a semantically equivalent DAG which it can execute more efficiently. It does this in two ways. (1)~\tool merges a chain of operations to avoid unnecessary work. For example, in \Cref{fig:design:dag}, calling \texttt{url-decode} before running \texttt{regex} is unnecessary and can be optimized. To this end, \tool removes the \texttt{url-decode} operation and rephrases the \texttt{regex} pattern. (2)~\tool merges variables, and consequently their downstream operations, to reduce the number of operation invocations. As we will show in \Cref{sec:eval}, this can improve the performance of some operations, e.g. \texttt{regex}. For example, in \Cref{fig:design:dag}, the downstream subgraph of the \texttt{user-agent} and \texttt{path} variable consists of the same operations, but might use a different regex pattern. \tool merges both variables, and adapts their subgraph, i.e. the regex pattern, accordingly.

\paragraph{Parallelizing Independent DAG Components} \tool leverages the inherent parallelism of the DAG to generate an efficient execution plan. For example, the DAG visualized in \Cref{fig:design:dag} consists of two independent components, allowing \tool to execute both in parallel.

\paragraph{Main Challenge} The generation of an optimal execution plan is an NP-hard problem. Both DAG optimizations depend on each other, and thus cannot be applied greedily. Moreover, the performance benefits heavily depend on the configured rule set and the ingress traffic. Thus, to approximate an efficient execution plan, emulating it prior to deployment, and monitoring it during operation, is necessary.

\section{Preliminary Evaluation}
\label{sec:eval}

We demonstrate the potential of \tool's DAG optimizations with two micro benchmarks that measure the relative speedup of merging operations, and variables, respectively. We compare both optimizations against ModSecurity's execution model, which sequentially executes each operation for every input variable. As input, we randomly select HTTP headers from the CSE-CIC-IDS2018~\cite{sharafaldin_csecicids2018}, the state-of-the-art intrustion detection system (IDS) dataset, and regex patterns from CRS~\cite{online_crs}. Each data point constitutes an average of a 1000 samples to reduce noise.

\begin{figure}[t]
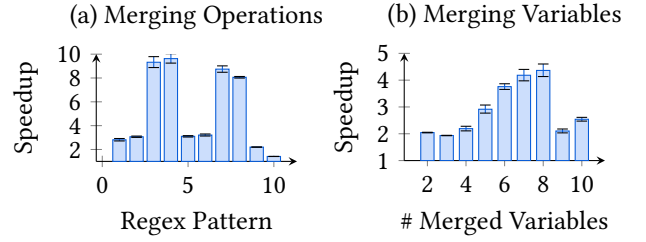

    \begin{subfigure}[t]{0.5\linewidth}
        \tikzsetnextfilename{merge-ops-figure}

\begin{tikzpicture}
\begin{axis}[
title={(a) Merging Operations},
xlabel={Regex Pattern},
ylabel={Speedup},
axis lines=left,
xmin=1,
xmax=10,
ymin=1,
ymax=10,
bar width=5pt,
enlarge x limits=0.15,
scaled x ticks=false,
scaled y ticks=false,
height=3cm,
width=\linewidth]

\input{fig/data/merge-ops}

\end{axis}
\end{tikzpicture}
    \end{subfigure}%
    \begin{subfigure}[t]{0.5\linewidth}
        \tikzsetnextfilename{merge-vars-figure}

\begin{tikzpicture}
\begin{axis}[
title={(b) Merging Variables},
xlabel={\# Merged Variables},
ylabel={Speedup},
axis lines=left,
xmin=2,
xmax=10,
ymin=1,
ymax=5,
bar width=5pt,
enlarge x limits=0.15,
scaled x ticks=false,
scaled y ticks=false,
height=3cm,
width=\linewidth]

\input{fig/data/merge-vars}

\end{axis}
\end{tikzpicture}
    \end{subfigure}%
    \caption{Merging DAG operations and variables improves the performance compared to the naive execution by at least $2\times$.}
    \label{fig:eval:merging}
\end{figure}

\paragraph{Merging Operations} We evaluate the speedup of merging a \texttt{url-decode} and \texttt{regex} operation. We select ten regex patterns at random, and report the relative speedup in comparison to a sequential execution. The result is visualized in \Cref{fig:eval:merging}a. We find that this optimization yields at least a $2\times$ speedup, and a $10\times$ speedup at most. Crucially, however, this optimization never degrades the performance.

\paragraph{Merging Variables} We measure the relative speedup of merging up to ten variables, each with a different downstream \texttt{regex} operation, and report the relative speedup in comparison to a sequential execution. The result is visualized in \Cref{fig:eval:merging}b. It shows that this optimization yields at least a $2\times$ speedup. We find that the speedup tends to increase linearly with the number of regex patterns, but does not scale at that rate beyond eight variables. Further analysis is needed to understand this scaling behavior.

\section{Conclusion and Future Work}

We presented \tool and demonstrated its potential to optimize WAFs. Future efforts will focus on the implementation of the DAG model and an end-to-end evaluation. Moreover, we believe that this architecture lends itself to further acceleration by offloading some rules to kernel space with eBPF. We intend to open-source \tool to foster reproducibility.

\bibliographystyle{acm}
\bibliography{bib/library.bib,bib/proj.bib}

\end{document}